\documentclass[letterpaper, 10 pt, conference]{ieeeconf}
\IEEEoverridecommandlockouts
\usepackage[utf8]{inputenc}
\usepackage{amsmath}
\usepackage{amssymb}
\usepackage{graphicx,balance}
\usepackage{pifont}
\usepackage{bm}
\usepackage{dsfont}
\usepackage{subcaption}
\usepackage{tikz}

\usepackage{amsfonts}
\usepackage{url}
\usepackage[pdftex,plainpages = false, colorlinks=true, linkcolor=black, citecolor = black, urlcolor = blue,pagebackref=false,hypertexnames=false, plainpages=false, pdfpagelabels]{hyperref}
\usepackage{balance}
\usepackage[capitalize]{cleveref}
\usepackage[sort,compress]{cite}

\newtheorem{theorem}{Theorem}[section]

\newtheorem{lemma}[theorem]{Lemma}

\crefname{section}{Section}{Sections}
\crefname{theorem}{Theorem}{Theorems}
\crefname{lemma}{Lemma}{Lemmas}
\crefname{table}{Table}{Tables}
\crefformat{equation}{(#2#1#3)}
\crefname{algocf}{Algorithm}{Algorithms}
\Crefname{algocf}{Algorithm}{Algorithms}
\crefname{ALC@unique}{Line}{Lines}

\newcommand{\x}{\mathbf{x}}

\newcommand{\z}{\mathbf{z}}

\newcommand{\R}{\mathds{R}}

\newcommand{\istate}{k}

\newcommand{\xnom}{\mathring{\mathbf{x}}}

\newcommand{\bpxnom}{\mathcal{B}_p(\xnom, \bm{\epsilon)}}

\newcommand{\CROWNAu}{\mathbf{\Psi}}
\newcommand{\CROWNAl}{\mathbf{\Phi}}
\newcommand{\CROWNbu}{\bm{\alpha}}
\newcommand{\CROWNbl}{\bm{\beta}}

\usepackage{accents}
\newcommand{\ubar}[1]{\underaccent{\bar}{#1}}

\newcommand{\nstep}{ReBReach-LP}
\newcommand{\basic}{BReach-LP}
\usepackage{paralist}

\usepackage[addedmarkup=bf]{changes}
\definechangesauthor[name={MFE}, color={blue}]{MFE}
\definechangesauthor[name={JH}, color={red}]{JH}
\definechangesauthor[name={NR}, color={orange}]{NR}

\crefformat{chapter}{\S#2#1#3}
\crefmultiformat{chapter}{\S\S#2#1#3}{and~#2#1#3}{, #2#1#3}{, and~#2#1#3}
\crefformat{section}{\S#2#1#3}
\crefmultiformat{section}{\S\S#2#1#3}{and~#2#1#3}{, #2#1#3}{, and~#2#1#3}

\usepackage{algorithm,algorithmic}

\title{\LARGE \bf
Backward Reachability Analysis for Neural Feedback Loops
}

\author{Nicholas Rober, Michael Everett, and Jonathan P. How %
\thanks{Aerospace Controls Laboratory, Massachusetts Institute of Technology, Cambridge, USA. e-mail: {\tt \small \{nrober,mfe,jhow\}@mit.edu}. Research supported by Ford Motor Company.}}

\begin{document}

\maketitle

\begin{abstract}
    The increasing prevalence of neural networks (NNs) in safety-critical applications calls for methods to certify their behavior and guarantee safety.
    This paper presents a backward reachability approach for safety verification of neural feedback loops (NFLs), i.e., closed-loop systems with NN control policies.
    While recent works have focused on forward reachability as a strategy for safety certification of NFLs, backward reachability offers advantages over the forward strategy, particularly in obstacle avoidance scenarios.
    Prior works have developed techniques for backward reachability analysis for systems without NNs, but the presence of NNs in the feedback loop presents a unique set of problems due to the nonlinearities in their activation functions and because NN models are generally not invertible.
    To overcome these challenges, we use existing forward NN analysis tools to find affine bounds on the control inputs and solve a series of linear programs (LPs) to efficiently find an approximation of the backprojection (BP) set, i.e., the set of states for which the NN control policy will drive the system to a given target set.
    We present an algorithm\footnote{\textbf{Code}: \url{https://github.com/mit-acl/nn_robustness_analysis}} to iteratively find BP set estimates over a given time horizon and demonstrate the ability to reduce conservativeness in the BP set estimates by up to 88\% with low additional computational cost.
    We use numerical results from a double integrator model to verify the efficacy of these algorithms and demonstrate the ability to certify safety for a linearized ground robot model in a collision avoidance scenario where forward reachability fails.
\end{abstract}


\section{Introduction}

Neural networks (NNs) play an important role in many modern robotic systems.
However, despite achieving high performance in nominal scenarios, many works have demonstrated that NNs can be sensitive to small perturbations in the input space \cite{kurakin2016adversarial, yuan2019adversarial}.
Thus, before applying NNs to safety-critical systems such as self-driving cars \cite{chen2015deepdriving} and aircraft collision avoidance \cite{julian2019deep}, there is a need for tools that provide safety guarantees, which presents computational challenges due to the high dimensionality and nonlinearities of NNs.

\begin{figure}[t]
\setlength\belowcaptionskip{-0.7\baselineskip}
\centering
\captionsetup[subfigure]{aboveskip=-1pt,belowskip=-1pt}
    \begin{subfigure}[t]{\columnwidth}
        \includegraphics[width=\columnwidth]{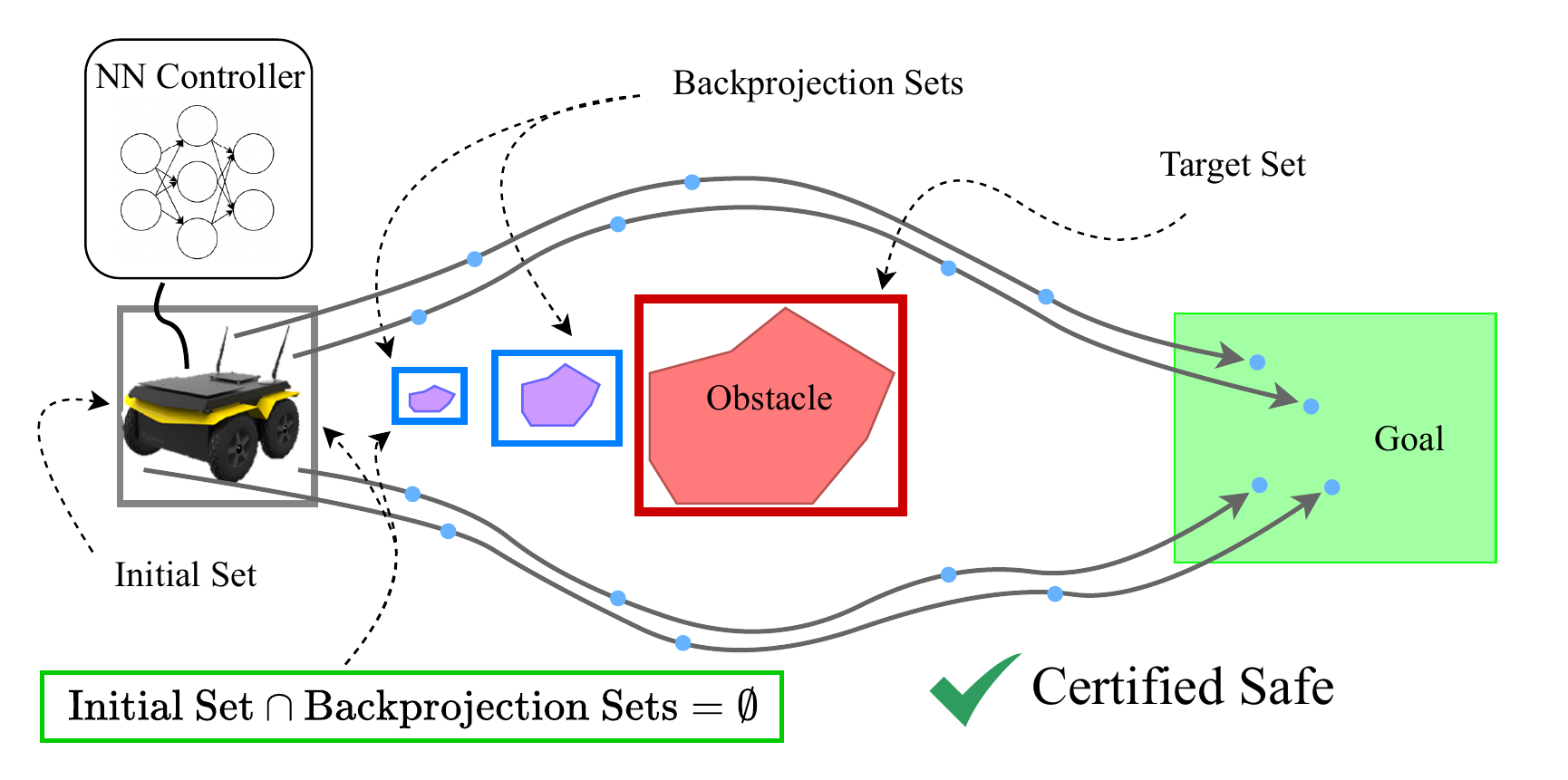}
        \caption{Backward reachability strategy for collision avoidance. The BP set estimates define the set of states that will lead to the obstacle, thus if the initial state set does not intersect with any BPs, the situation is safe.}
        \label{fig:reach_comp:backward}
    \end{subfigure}
    \begin{subfigure}[t]{\columnwidth}
        \includegraphics[width=\columnwidth]{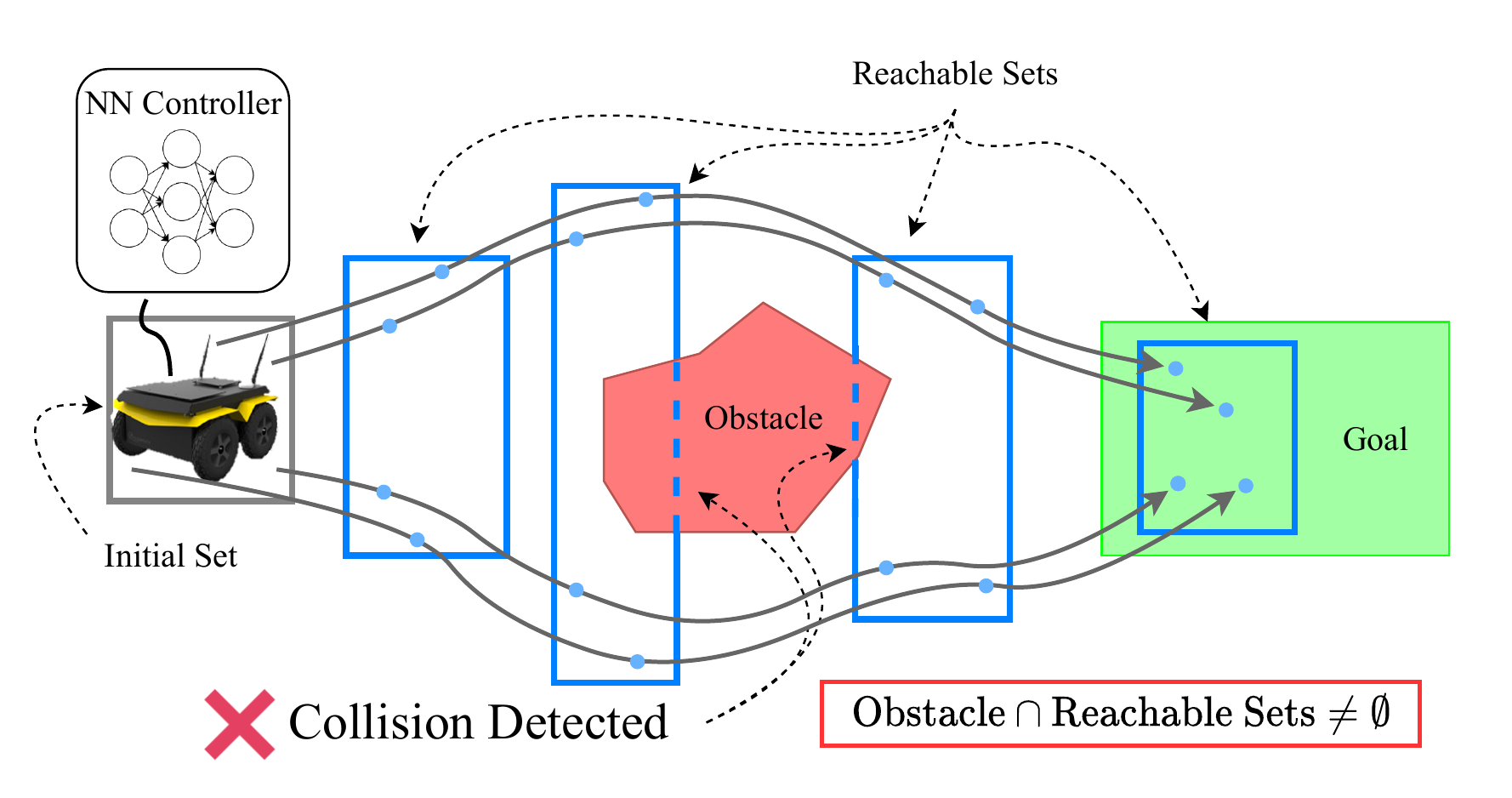}
        \caption{Forward reachability strategy for collision avoidance. The reachable set estimates define the set of possible future states the system will be in, thus any intersection of an obstacle means safety cannot be certified.}
        \label{fig:reach_comp:forward}
    \end{subfigure}
    \caption{Collision-avoidance scenario where backward reachability is able to correctly guarantee safety whereas forward reachability fails.}
    \label{fig:reach_comp}
\end{figure}

Numerous tools have recently been developed to analyze both NNs in isolation \cite{zhang2018efficient, weng2018towards, xu2020automatic, tjeng2017evaluating, katz2019marabou, katz2017reluplex, vincent2021reachable, jia2021verifying} and neural feedback loops (NFLs), e.g., closed-loop systems with NN control policies, \cite{dutta2019reachability, huang2019reachnn, ivanov2019verisig, fan2020reachnn, hu2020reach, sidrane2021overt, everett2021reachability, bak2022closed}. 
While many of these tools focus on forward reachability \cite{dutta2019reachability, huang2019reachnn, ivanov2019verisig, fan2020reachnn, hu2020reach, sidrane2021overt, everett2021reachability}, which certifies safety by estimating where the NN will drive the system, this work focuses on backward reachability \cite{bak2022closed}, as shown in \cref{fig:reach_comp:backward}.
Backward reachability  accomplishes safety certification by finding \textit{backprojection} (BP) \textit{sets} that define parts of the state space for which the NN will drive the system to the target set, which can be chosen to contain an obstacle.
Backward reachability offers an advantage over forward reachability in scenarios where the possible future trajectories diverge in multiple directions.
This phenomenon is demonstrated by the collision-avoidance scenario in \cref{fig:reach_comp:forward} where forward reachability is used and the robot's position within the initial state set determines whether the vehicle will go above or below the obstacle.
When using a single convex representation of reachable sets, forward reachability analysis will be unable to certify safety because the reachable set estimates span the two sets of possible trajectories, thus intersecting with the obstacle.
Conversely, as shown in \cref{fig:reach_comp:backward}, backward reachability analysis correctly evaluates the situation as safe because the vehicle starts outside the avoid set's BP, and thus the vehicle is guaranteed to avoid the obstacle.
Moreover, in the ideal case that the NN control policy is always able to avoid an obstacle, the true BP set will be empty, allowing the algorithm to terminate, thereby reducing the computational cost compared to a forward reachability strategy that must calculate reachable sets for the full time horizon.



While forward and backward reachability differ only by a change of variables for systems without NNs~\cite{bansal2017hamilton,evans1998partial,mitchell2007comparing}, both the nonlinearities and dimensions of the matrices associated with NN controllers lead to fundamental challenges that make propagating sets backward through an NFL complicated.
Despite promising prior work \cite{vincent2021reachable,bak2022closed,everett2021reachability}, there are no existing techniques that efficiently find BP set estimates over multiple timesteps for the general class of linear NFLs considered in this work. Our work addresses this issue by using a series of NN relaxations to constrain a set of linear programs (LPs) that can be used to find BP set approximations that are guaranteed to contain the true BP set.
We leverage CROWN \cite{zhang2018efficient}, an efficient open-loop NN verification tool, to generate affine bounds on the NN output for a given set of inputs. These bounds are used to constrain the system input and solve an LP maximizing the size of the BP set subject to constraints on the dynamics and control limits. The contributions of this work include:

\begin{compactitem}
    \item \basic{}: an LP-based technique to efficiently find multi-step BP over-approximations for NFLs that can be used to guarantee that the system will avoid collisions,
    \item \nstep{}: an algorithm to refine multiple one-step BP over-approximations, reducing conservativeness in the BP estimate by up to 88\%, 
    \item Numerical experiments that exhibit our BP estimation techniques with a double integrator model and a demonstration certifying safety for a linearized ground robot model whereas the forward reachability tools proposed in \cite{everett2021reachability,sidrane2021overt} fail.
\end{compactitem}




\section{Related Work}
Reachability analysis can be broadly categorized into three categories by system type: NNs in isolation (i.e., open-loop analysis), closed-loop systems without neural components, and NFLs.

Open-loop NN analysis encompasses techniques that relax the nonlinearities in the NN activation functions to quickly provide relatively conservative bounds on NN outputs~\cite{zhang2018efficient, raghunathan2018semidefinite}, and techniques that take more time to provide exact bounds~\cite{katz2017reluplex, vincent2021reachable}.
Many of these open-loop studies are motivated by the goal of guaranteeing robustness to adversarial attacks against perception models~\cite{raghunathan2018semidefinite, tjeng2017evaluating}, but cannot be directly used to guarantee safety for closed-loop systems because they do not consider closed-loop system dynamics.


For closed-loop systems without NNs, reachability analysis is a well established method of providing safety verification.
Hamilton-Jacobi methods~\cite{bansal2017hamilton, evans1998partial}, CORA~\cite{althoff2015introduction}, Flow*~\cite{chen2013flow}, SpaceEx~\cite{frehse2011spaceex}, and C2E2~\cite{duggirala2015c2e2} are tools that are commonly used for reachability analysis, but because they do not handle open loop NN analysis, they cannot be used to analyze NFLs.


Forward reachability analysis is the focus of many recent works~\cite{dutta2019reachability, huang2019reachnn, ivanov2019verisig, fan2020reachnn, hu2020reach, sidrane2021overt, everett2021reachability}, but while more traditional approaches to reachability analysis, e.g., Hamilton-Jacobi methods, can easily switch from forward to backward with a change of variables~\cite{bansal2017hamilton, evans1998partial, xue2016under, kochdumper2020computing, yang2021scalable}, backward reachability for NFLs is less straightforward.
One challenge with propagating sets backward through a NN is that many activation functions have finite range, meaning that there is not a one-to-one mapping of inputs to outputs (e.g., $\mathrm{ReLU}(x) = 0$ corresponds to all values of $x \leq 0$), which can cause large amounts of conservativeness in the BP set estimate as there is an infinite set of possible inputs.
Additionally, even if an infinite-range activation function is used, NN weight matrices may be singular or rank deficient and are thus generally not invertible, again causing problems with determining inputs given a set of outputs.
While recent works on NN inversion have developed NN architectures that are designed to be invertible~\cite{ardizzone2018analyzing} and training procedures that regularize for invertibility~\cite{behrmann2019invertible}, dependence on these techniques would be a major limitation on the class of systems for which backward reachability analysis could be applied.
Our approach avoids the challenges associated with finite-range activation functions and NN-invertibility and can be applied to the same class of NN architectures as CROWN \cite{zhang2018efficient}, i.e. NNs for which an affine relaxation can be found.

Several recent works have investigated backwards reachability analysis for NFLs.
Ref.~\cite{vincent2021reachable} describes a method for open-loop backward reachability on a NN dynamics model, but this method does not directly apply to this work's problem of interest, namely a NN controller with known dynamics.
Alternatively, while~\cite{bak2022closed} analyzes NFLs, they use a quantized state approach~\cite{jia2021verifying} that requires an alteration of the original NN through a preprocessing step that can affect its overall behavior.
Finally, previous work by the authors~\cite{everett2021reachability} derives a closed-form equation that can be used to find under-approximations of the BP set, but this is most useful for goal checking when it is desirable to guarantee that all states in the BP estimate will reach the target set.
This work builds off of~\cite{everett2021reachability}, adapting some of the steps used to find BP under-approximations to instead find BP over-approximations, which are good for obstacle avoidance because they contain all the states that reach the target set.

\section{Preliminaries}

\subsection{System Dynamics}

We first assume that the system of interest can be described by the linear discrete-time system,
\begin{align}
\begin{split}
    \mathbf{x}_{t+1} & = \mathbf{A} \mathbf{x}_{t} + \mathbf{B} \mathbf{u}_t + \mathbf{c} \label{eqn:ltv_dynamics} \\
    \mathbf{y}_{t} & = \mathbf{C}^T\mathbf{x}_t,
\end{split}
\end{align}
where $\mathbf{x}_t\in\R^{n_x},\, \mathbf{u}_t\in\R^{n_u},\, \mathbf{y}_t\in\R^{n_y}$ are state, control, and output vectors, $\mathbf{A},\, \mathbf{B},\, \mathbf{C}$ are known system matrices, and $\mathbf{c}\in\R^{n_x}$ is a known exogenous input.
We assume the control input is constrained by control limits, i.e., $\mathbf{u}_t\in\mathcal{U}$, and is determined by a state-feedback control policy $\mathbf{u}_t=\pi(\mathbf{x}_t)$ (i.e., $\mathbf{C}=\mathbf{I}_{n_x}$) where $\pi(\cdot)$ is an $m$-layer feedforward NN.
Denote the closed-loop system \cref{eqn:ltv_dynamics} and control policy $\pi$ as 
\begin{equation}
    \mathbf{x}_{t+1} = f(\mathbf{x}_{t}; \pi). \label{eqn:closed_loop_dynamics}
\end{equation}

\subsection{Control Policy Neural Network Structure}


Consider a feedforward NN with $L$ hidden layers and two additional layers for input and output.
We denote the number of neurons in each layer as $n_l\ \forall l \in [L+1]$ where $[i]$ denotes the set $\{0,1,\ldots,i\}$.
The $l$-th layer has weight matrix $\mathbf{W}^{l}\in\R^{n_{l+1}\times n_{l}}$, bias vector $\mathbf{b}^{l}\in\R^{n_{l+1}}$, and activation function $\sigma^{l}: \R^{n_{l+1}} \to \R^{n_{l+1}}$, where $\sigma^{l}$ can be any option handled by CROWN~\cite{zhang2018efficient}, e.g., sigmoid, tanh, ReLU, etc. For an input $\x\in\R^{n_0}$, the NN output $\pi(\x)$ is computed as
\begin{align}
\begin{split}
    \x^{0} &= \x \\
    \z^{l} &= \mathbf{W}^{l} \x^{l}+\mathbf{b}^{l}, \forall l\in[L] \\
    \x^{l+1} &= \sigma^{l}(\z^{l}), \forall l\in[L-1] \\
    \pi(\x) &= \z^{L}.
\end{split}
\end{align}

\subsection{Neural Network Robustness Verification}



To avoid the computational cost associated with calculating exact BP sets, we relax the NN's activation functions to obtain affine bounds on the NN outputs for a known set of inputs.
The range of inputs are represented using the $\ell_p$-ball 
\begin{align}
    \bpxnom &\triangleq \{\mathbf{x}\ \lvert\ \lim_{\bm{\epsilon}' \to \bm{\epsilon}^+} \lvert\lvert (\mathbf{x} - \xnom) \oslash \bm{\epsilon}' \rvert\rvert_p \leq 1\},
\end{align}
where $\xnom \in \R^n$ is the center of the ball, $\bm{\epsilon}\in\R^n_{\geq 0}$ is a vector whose elements are the radii for the corresponding elements of $\mathbf{x}$, and $\oslash$ denotes element-wise division.

\begin{theorem}[\!\!\cite{zhang2018efficient}, Convex Relaxation of NN]\label{thm:crown_particular_x}
Given an $m$-layer neural network control policy $\pi:\R^{n_x}\to\R^{n_u}$, there exist two explicit functions $\pi_j^L: \R^{n_x}\to\R^{n_u}$ and $\pi_j^U: \R^{n_x}\to\R^{n_u}$ such that $\forall j\in [n_m], \forall \mathbf{x}\in\bpxnom$, the inequality $\pi_j^L(\mathbf{x})\leq \pi_j(\mathbf{x})\leq \pi_j^U(\mathbf{x})$ holds true, where
\begin{equation}
\label{eq:f_j_UL}
    \pi_{j}^{U}(\x) = \CROWNAu_{j,:} \x + \CROWNbu_j, \quad
    \pi_{j}^{L}(\x) = \CROWNAl_{j,:} \x + \CROWNbl_j,
\end{equation}
where $\CROWNAu, \CROWNAl \in \R^{n_u \times n_x}$ and $\CROWNbu, \CROWNbl \in \R^{n_u}$ are defined recursively using NN weights, biases, and activations (e.g., ReLU, sigmoid, tanh), as detailed in~\cite{zhang2018efficient}.
\end{theorem}



\subsection{Backreachable \& Backprojection Sets}

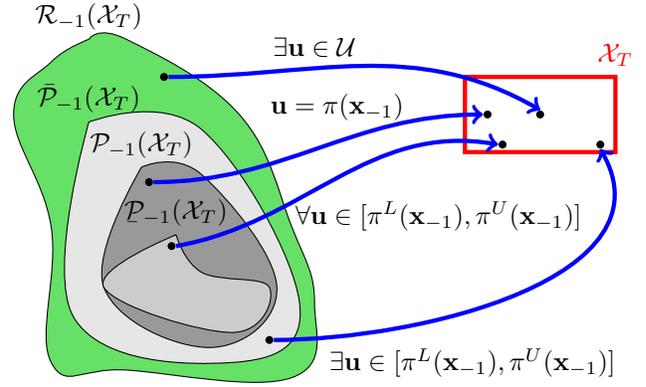
\begin{figure}[t]
\begin{tikzpicture}

    \draw [red, ultra thick] (2,1) rectangle (4,2) node[above] {$\mathcal{X}_{T}$};
    
    \draw [fill=black!20!green!60!white]  plot[smooth, tension=.7] coordinates {(-3,2.5) (-3.5,2.2) (-4,1.5) (-3.5,0) (-3.5,-1.8) (-2.5,-1.9) (-0.3,-2) (0.,-1) (-0.3,0) (-1,1.5) (-2.5,2.5) (-3,2.5)} node[above] {$\mathcal{R}_{-1}(\mathcal{X}_T)$};
    \node[circle,fill=black,inner sep=0pt,minimum size=3pt] (a) at (-2,2) {};
    \node[circle,fill=black,inner sep=0pt,minimum size=3pt] (b) at (3,1.5) {};
    \draw [->,blue, ultra thick] (a.east) to [out=0,in=150] (b.north);
    \node[black] (RTset) at (-0,2.4) {$\exists \mathbf{u} \in \mathcal{U}$};

    \draw [fill=black!10!white]  plot[smooth, tension=.7] coordinates {(-3.0,1.4) (-1.5,1.3) (-0.2,-1.0) (-1.1,-1.8) (-3.2,-1.2) (-3.0,1.4)} node[above] {$\bar{\mathcal{P}}_{-1}(\mathcal{X}_T)$};
    \node[circle,fill=black,inner sep=0pt,minimum size=3pt] (c) at (-0.6,-1.5) {};
    \node[circle,fill=black,inner sep=0pt,minimum size=3pt] (d) at (3.8,1.1) {};
    \draw [->,blue, ultra thick] (c.east) to [out=0,in=300] (d.south);
    \node[black] (PTbarset) at (2.1, -1.8) {$\exists \mathbf{u} \in [\pi^L(\mathbf{x}_{-1}), \pi^U(\mathbf{x}_{-1})]$};

    \draw [fill=black!40!white]  plot[smooth, tension=.7] coordinates {(-2.3,0.8) (-1.5,0.7) (-0.5,-0.8) (-1.4,-1.6) (-2.8,-0.8) (-2.3,0.8)} node[above] {$\mathcal{P}_{-1}(\mathcal{X}_T)$};
    \node[circle,fill=black,inner sep=0pt,minimum size=3pt] (e) at (-2.2,0.6) {};
    \node[circle,fill=black,inner sep=0pt,minimum size=3pt] (f) at (2.3,1.5) {};
    \draw [->,blue, ultra thick] (e.east) to [out=0,in=180] (f.west);
    \node[black] (PTset) at (0.3, 1.6) {$\mathbf{u} = \pi(\mathbf{x}_{-1})$};

    \draw [fill=black!20!white]  plot[smooth, tension=.7] coordinates {(-1.84,-0.1) (-1.5,-0.6) (-0.6,-0.8) (-1.1,-1.4) (-2.8,-0.8) (-1.84,-0.1)} node[above] {$\ubar{\mathcal{P}}_{-1}(\mathcal{X}_T)$};
    \node[circle,fill=black,inner sep=0pt,minimum size=3pt] (c) at (-1.9,-.25) {};
    \node[circle,fill=black,inner sep=0pt,minimum size=3pt] (d) at (2.5,1.1) {};
    \draw [->,blue, ultra thick] (c.east) to [out=10,in=170] (d.west);
    \node[black] (PTbarset) at (1.65, 0.15) {$\forall \mathbf{u} \in [\pi^L(\mathbf{x}_{-1}), \pi^U(\mathbf{x}_{-1})]$};

\end{tikzpicture}
\caption{
Backreachable, backprojection, and target sets.
Given a target set, $\mathcal{X}_T$, the backreachable set $\mathcal{R}_{-1}(\mathcal{X}_T)$ contains all states for which \textit{some} control exists to move the system to $\mathcal{X}_T$ in one timestep.
BP set $\mathcal{P}_{-1}(\mathcal{X}_T)$ contains all states for which the NN controller leads the system to $\mathcal{X}_T$.
BP under-approximation $\protect\ubar{\mathcal{P}}_{-1}(\mathcal{X}_T)$ and over-approximation $\bar{\mathcal{P}}_{-1}(\mathcal{X}_T)$ contain states for which \textit{all} and \textit{some}, respectively, controls that the relaxed NN could apply lead the system to $\mathcal{X}_T$.
}
\label{fig:backreachable}
\vspace*{-0.12in}
\end{figure}

The distinction between sets used in this work is shown in \cref{fig:backreachable}.
Given a convex target set $\mathcal{X}_T$ (right), each of the four sets on the left contain states that will reach $\mathcal{X}_T$ under different conditions on the control input $\mathbf{u}$, described below.

First, the one-step backreachable set
\begin{equation}
    \mathcal{R}_{-1}(\mathcal{X}_T) \triangleq \{ \mathbf{x}\ \lvert\ \exists\mathbf{u} \in \mathcal{U} \mathrm{\ s.t.\ } \label{eqn:backreachable_sets}
     \mathbf{A}\mathbf{x} + \mathbf{B}\mathbf{u} + \mathbf{c} \in \mathcal{X}_T \},
\end{equation}
contains the set of all states that transition to $\mathcal{X}_T$ in one timestep given some $\mathbf{u}\in \mathcal{U}$.
The importance of the backreachable set (or at least an over-approximation to the backreachable set, as will be seen later) is that it only depends on the control limits $\mathcal{U}$ and not the NN control policy $\pi$.
Thus, while $\mathcal{R}_{-1}(\mathcal{X}_T)$ is itself a very conservative over-approximation of the true set of states that will reach $\mathcal{X}_T$ under $\pi$, it provides a region over which we can relax the NN with forward NN analysis tools, thereby avoiding issues with NN invertibility.

Next, we define the one-step true BP set as
\begin{equation}
    \mathcal{P}_{-1}(\mathcal{X}_T) \triangleq \{ \mathbf{x}\ \lvert\ \mathbf{A}\mathbf{x} + \mathbf{B}\pi(\mathbf{x}) + \mathbf{c} \in \mathcal{X}_T \}, \label{eqn:backprojection_sets}
\end{equation}
which denotes the set of all states that will reach $\mathcal{X}_T$ in one timestep given the input from $\pi$.
As previously noted, calculating $\mathcal{P}_{-1}(\mathcal{X}_T)$ exactly is computationally intractable, which motivates the development of approximation techniques.
Thus, the final two sets shown in~\cref{fig:backreachable} are the BP set over-approximation 
\begin{align}
    \bar{\mathcal{P}}_{-1}(\mathcal{X}_T) &\triangleq \{ \mathbf{x}\ \lvert\ \exists\mathbf{u} \in [\pi^L(\x), \pi^U(\x)] \mathrm{\ s.t.\ } \label{eqn:backprojection_set_over} \\ \nonumber
     &\mathbf{A}\mathbf{x} + \mathbf{B}\mathbf{u} + \mathbf{c} \in \mathcal{X}_T \},\
\end{align}
and BP set under-approximation
\begin{align}
    \ubar{\mathcal{P}}_{-1}(\mathcal{X}_T) &\triangleq \{ \mathbf{x}\ \lvert\ \forall\mathbf{u} \in [\pi^L(\x), \pi^U(\x)] \mathrm{\ s.t.\ } \label{eqn:backprojection_set_under} \\ \nonumber
     &\mathbf{A}\mathbf{x} + \mathbf{B}\mathbf{u} + \mathbf{c} \in \mathcal{X}_T \}.\
\end{align}
Comparison of the motivation behind over- and under-approximation strategies is given in the next section.

\subsection{Backprojection Set: Over- vs. Under-Approximations}

The need to approximate the BP set naturally leads to the question of whether we should compute over- or under-approximations.
Both types of BP set approximations have relevant physical meaning and are valuable for different reasons.
An under-approximation is useful if the target set is a goal set, because we aim to find a set of states at the previous timestep that will \textit{certainly} drive the system into the goal set.
This leads to a ``for all'' condition on the relaxed NN (i.e., $\forall \mathbf{u} \in [\pi^L(\x), \pi^U(\x)]$) in \cref{eqn:backprojection_set_under}. Conversely, an over-approximation is useful if the target set is an obstacle/avoid set, because we aim to find all states at the previous timestep for which $\pi$ \textit{could} drive the system into the avoid set.
This leads to an ``exists'' condition on the relaxed NN (i.e., $\exists \mathbf{u} \in [\pi^L(\x), \pi^U(\x)]$) in \cref{eqn:backprojection_set_over}. Note that in this work we use ``over-approximation'' and ``outer-bound'' interchangeably.

Ref.~\cite{everett2021reachability} introduced a closed-form equation capable of one-step under-approximations of BP sets.
Unfortunately, this closed form equation hinged on the ``for all'' condition and thus cannot be used to generate BP set over-approximations, necessitating a different approach. 


\section{Approach}
This section first outlines a technique to find one-step BP set over-approximations (\cref{alg:one_step_backprojection}) by solving a series of LPs. 
We then introduce \basic{}, which iteratively calls \cref{alg:one_step_backprojection} to calculate BP set estimates over a desired time horizon.
Finally, we propose \nstep{}, which further refines the BP set estimates from \basic{} with another series of LPs, thus reducing conservativeness with some additional computational cost.

\subsection{Over-Approximation of 1-Step Backprojection Sets} 
\label{sec:overapprox_one_step_BP}

The proposed approach is as follows:
\begin{enumerate}
    \item Ignoring the NN and using control limits $\mathcal{U}$, solve two LPs for each element of the state vector to find the hyper-rectangular bounds $\bar{\mathcal{R}}_{-1}$ on the backreachable set $\mathcal{R}_{-1}$ (note that $\bar{\mathcal{R}}_{-1} \supseteq \mathcal{P}_{-1}$)
    \item Find upper/lower affine control bounds $\pi^U(\x_t)$ and $\pi^L(\x_t)$ by relaxing the NN controller (we use CROWN \cite{zhang2018efficient}, but other tools, e.g., \cite{weng2018towards, xu2020automatic}, could also be used) within the backreachable set
    \label{enum:nn_relaxation}
    \item Solve two LPs for each element of the state vector to compute hyper-rectangular bounds $\bar{\mathcal{P}}_{-1}$ on the states that will lead to the target set for \textit{some} control effort within the upper/lower bounds calculated in Step \ref{enum:nn_relaxation}
\end{enumerate}
The last step gives an over-approximation of the BP set, which is the set of interest.

The following lemma provides hyper-rectangle bounds on $\bar{\mathcal{P}}_{t}(\mathcal{X}_T)$ for a single timestep, which is the key component of the recursive algorithm introduced in the next section.

\begin{lemma}\label[lemma]{thm:backprojection}

Given an $m$-layer NN control policy $\pi:\R^{n_y}\to\R^{n_u}$, closed-loop dynamics $f: \R^{n_x} \times \Pi \to \R^{n_x}$ as in~\cref{eqn:ltv_dynamics,eqn:closed_loop_dynamics}, and target set $\mathcal{X}_T$, the set 
\begin{equation}
    \bar{\mathcal{P}}_{-1}(\mathcal{X}_T) = \{\x_t\ |\ \ubar{\x}_t \leq \x_t \leq \bar{\x}_t \}
\end{equation}
is a superset of the true BP set $\mathcal{P}_{-1}(\mathcal{X}_T)$,
where $\ubar{\mathbf{x}}_t$ and $\bar{\mathbf{x}}_t$ are computed elementwise by solving the LPs in~\cref{eqn:backprojection:lp_obj}.

\begin{proof}
Given dynamics~\cref{eqn:ltv_dynamics,eqn:closed_loop_dynamics} and constraints 
\begin{equation}
\mathcal{F_{\bar{R}}} \triangleq \left\{
\mathbf{x}_t, \mathbf{u}_t\ \left| \
\begin{aligned}
& \mathbf{A} \mathbf{x}_t + \mathbf{B} \mathbf{u}_t + \mathbf{c} \in \mathcal{X}_T,\\
& \mathbf{u}_t \in \mathcal{U},
\end{aligned}\right.\kern-\nulldelimiterspace
\right\},
\label{eqn:backreachable:lp_constr}
\end{equation}
solve these optimization problems for each state $\istate \in [n_x]$,
\begin{equation}
\bar{\bar{\mathbf{x}}}_{t; \istate} = \min_{\mathbf{x}_t, \mathbf{u}_t \in \mathcal{F_{\bar{R}}}} \mathbf{e}_k^\top  \mathbf{x}_{t}, \quad \quad \ubar{\ubar{\mathbf{x}}}_{t; \istate} = \max_{\mathbf{x}_t, \mathbf{u}_t \in \mathcal{F_{\bar{R}}}} \mathbf{e}_k^\top  \mathbf{x}_{t}, \label{eqn:backreachable:lp_obj}
\end{equation}
where the notation $\mathbf{x}_{t;k}$ denotes the $k^\mathrm{th}$ element of $\mathbf{x}_t$ and $\mathbf{e}_k \ \in \R^{n_x}$ denotes the indicator vector, i.e., the vector with $k^\mathrm{th}$ element equal to one and all other elements equal to zero. Eq. \cref{eqn:backreachable:lp_obj} provides a hyper-rectangular outer bound $\bar{\mathcal{R}}_{-1}(\mathcal{X}_T) \triangleq \{\mathbf{x}\ | \ \ubar{\ubar{\mathbf{x}}}_{t} \leq \mathbf{x}_{t} \leq \bar{\bar{\mathbf{x}}}_{t}\}$ on the backreachable set.
Note that this is a LP for the convex $\mathcal{X}_T,\ \mathcal{U}$ used here.

Given the bound, $\ubar{\ubar{\mathbf{x}}}_{t} \leq \mathbf{x}_{t} \leq \bar{\bar{\mathbf{x}}}_{t}$, \cref{thm:crown_particular_x} provides $\CROWNAu, \CROWNAl, \CROWNbu, \CROWNbl$.
Then, define the set of state and control constraints $\mathcal{F_{\bar{P}}}$ as
\begin{equation}
\mathcal{F_{\bar{P}}} \triangleq \left\{
\mathbf{x}_t, \mathbf{u}_t\ \left| \
\begin{aligned}
& \mathbf{A} \mathbf{x}_t + \mathbf{B} \mathbf{u}_t + \mathbf{c} \in \mathcal{X}_T,\\
& \pi^L(\x_t) \leq \mathbf{u}_t \leq \pi^U(\x_t), \\
& \x_t \in \bar{\mathcal{R}}_{-1},
\end{aligned}\right.\kern-\nulldelimiterspace
\right\}
\label{eqn:backprojection:lp_constr}
\end{equation}
and solve the following optimization problems for each state $\istate \in [n_x]$:
\begin{equation}
\bar{\mathbf{x}}_{t; \istate} = \min_{\mathbf{x}_t, \mathbf{u}_t \in \mathcal{F_{\bar{P}}}} \mathbf{e}_k^\top \mathbf{x}_{t}, \quad \quad \ubar{\mathbf{x}}_{t; \istate} = \max_{\mathbf{x}_t, \mathbf{u}_t \in \mathcal{F_{\bar{P}}}} \mathbf{e}_k^\top \mathbf{x}_{t}. \label{eqn:backprojection:lp_obj}
\end{equation}


The LPs solved in  \cref{eqn:backprojection:lp_obj} provide a hyper-rectangular outer bound of the BP set, i.e.,
\begin{align}
    \bar{\mathcal{P}}_{-1}(\mathcal{X}_T) = &\ \{\x_t\ |\ \ubar{\x}_t \leq \x_t \leq \bar{\x}_t \}
    \label{eqn:BPproof:rect_bound} \\ 
    \begin{split}
         \supseteq & \ \{\x_t\ |\ \mathbf{A} \x_t + \mathbf{B} \mathbf{u}_t + \mathbf{c} \in \mathcal{X}_T,\ \\ & \quad \quad \quad \quad \pi^L(\x_t) \leq \mathbf{u}_t \leq \pi^U(\x_t)\},
    \end{split} \label{eqn:BPproof:true_relaxed_BP} \\
    \supseteq &\ \mathcal{P}_{-1}(\mathcal{X}_T).
    \label{eqn:BPproof:true_BP}
\end{align}
$\bar{\mathcal{P}}_{-1}(\mathcal{X}_T)$ is the hyper-rectangular outer bound of \cref{eqn:BPproof:true_relaxed_BP}, thus guaranteeing the relationship between \cref{eqn:BPproof:rect_bound} and \cref{eqn:BPproof:true_relaxed_BP}.
The relation between \cref{eqn:BPproof:true_relaxed_BP} and \cref{eqn:BPproof:true_BP} holds because $\pi^L(\x_t) \leq \pi(\x_t) \leq \pi^U(\x_t)$ (via \cref{thm:crown_particular_x}). 
It follows that $\bar{\mathcal{P}}_{-1}(\mathcal{X}_T) \supseteq \mathcal{P}_{-1}(\mathcal{X}_T)$.
\end{proof}
\end{lemma}

Note that we relaxed the NN over the entire range of $\bar{\mathcal{R}}_{-1}$, but if we instead divide it into smaller regions and relax each of them individually, we may find tighter bounds on the control input.
While partitioning $\bar{\mathcal{R}}_{-1}$ in this way is not required, it can be used to reduce the conservativeness of the BP estimate at the cost of increased computation time associated with solving LPs for each partitioned region.
Ref.~\cite{everett2021reachability} provides a detailed discussion on different partitioning strategies.




\subsection{Algorithm for Computing Backprojection Sets}

\begin{algorithm}[t]
 \caption{oneStepBackproj}
 \begin{algorithmic}[1]
 \setcounter{ALC@unique}{0}
 \renewcommand{\algorithmicrequire}{\textbf{Input:}}
 \renewcommand{\algorithmicensure}{\textbf{Output:}}
 \REQUIRE target state set $\mathcal{X}_T$, trained NN control policy $\pi$, partition parameter $\mathbf{r}$
 \ENSURE BP set approximation $\bar{\mathcal{P}}_{-1}(\mathcal{X}_T)$
    \STATE $\bar{\mathcal{P}}_{-1}(\mathcal{X}_T) \leftarrow \emptyset$
    \STATE $\bar{\mathcal{R}}_{t}(\mathcal{X}_T)\! =\! [\ubar{\ubar{\mathbf{x}}}_{t}, \bar{\bar{\mathbf{x}}}_{t}]\! \leftarrow\! \mathrm{backreach}(\mathcal{X}_T, \mathcal{U})$ 
    \STATE $\mathcal{S} \leftarrow \mathrm{partition}([\ubar{\ubar{\mathbf{x}}}_{t}, \bar{\bar{\mathbf{x}}}_{t}], \mathbf{r})$
    \FOR{$[\ubar{\mathbf{x}}_{t}, \bar{\mathbf{x}}_{t}]$ in $\mathcal{S}$}
        \STATE $\CROWNAu, \CROWNAl, \CROWNbu, \CROWNbl \leftarrow \mathrm{CROWN}(\pi, [\ubar{\mathbf{x}}_{t}, \bar{\mathbf{x}}_{t}])$ 
        \FOR{$\istate \in n_x$}
            \STATE $\bar{\bar{\mathbf{x}}}_{t;k} \leftarrow \mathrm{lpMax}(\mathcal{X}_T,\bar{\mathcal{R}}_t, \CROWNAu, \CROWNAl, \CROWNbu, \CROWNbl)$ 
            \STATE $\ubar{\ubar{\mathbf{x}}}_{t;k} \leftarrow \mathrm{lpMin}(\mathcal{X}_T,\bar{\mathcal{R}}_t,\CROWNAu, \CROWNAl, \CROWNbu, \CROWNbl)$ 
        \ENDFOR
        \STATE $\mathcal{A} \leftarrow \{\mathbf{x}\ \lvert\ \forall \istate \in n_x,\ \ubar{\ubar{\mathbf{x}}}_{t;k} \leq \mathbf{x} \leq \bar{\bar{\mathbf{x}}}_{t;k}\}$
        \STATE $\bar{\mathcal{P}}_{-1}(\mathcal{X}_T) \leftarrow \bar{\mathcal{P}}_{-1}(\mathcal{X}_T) \cup \mathcal{A}$ \label{alg:one_step_backprojection:union}
    \ENDFOR
 \RETURN $\bar{\mathcal{P}}_{-1}(\mathcal{X}_T)$ 
 \end{algorithmic}\label{alg:one_step_backprojection}
\end{algorithm}

\cref{alg:one_step_backprojection} follows the procedure outlined in \cref{sec:overapprox_one_step_BP} and \cref{thm:backprojection} to obtain $\bar{\mathcal{P}}_{-1}(\mathcal{X}_T)$, i.e., an over-approximation of the BP set for a single timestep.
The functions {\tt lpMax} and {\tt lpMin} solve the LPs formulated by \cref{eqn:backprojection:lp_obj} and {\tt backreach} solves the LPs formulated by \cref{eqn:backreachable:lp_obj}.
The partition parameter $\mathbf{r} \in \mathds{R}^{n_x}$ gives the option to uniformly split the backreachable set, e.g., $\mathbf{r} = [2,3]$ will split $\bar{\mathcal{R}}_{-1}(\mathcal{X}_T)$ into a $2 \times 3$ grid of cells.

To provide safety guarantees over an extended time horizon $\tau$, we extend this idea to iteratively compute BPs at multiple timesteps $\bar{\mathcal{P}}_{-\tau:0}(\mathcal{X}_T)$. 
We first initialize the zeroth BP set as the target set (\cref{alg:backprojection:initialize}). 
Then we step backward in time (\cref{alg:backprojection:timestep_for_loop}), recursively using \cref{alg:one_step_backprojection} ({\tt oneStepBackproj}), and the BP from the previous step to iteratively compute the new BP set (\cref{alg:backprojection:one_step_backproj_iter}). 
This is done $\tau$ times to give a list of BP set estimates $\bar{\mathcal{P}}_{-\tau:0}(\mathcal{X}_T)$.
The proposed procedure is summarized in \cref{alg:backprojection}.
Note that from this, we can see that the number of LPs solved $N_{LP}$ can be written as $N_{LP} = 2 n_x N_\mathbf{r} \tau$, where $N_\mathbf{r}$ is the number of partitions associated with $\mathbf{r}$ at each step.
Thus, the computational complexity is linear with respect to state dimension. 
However, $N_\mathbf{r}$ can grow quickly with state dimension, therefore necessitating future investigation into efficient partitioning strategies to reduce computation time.

\begin{algorithm}[t]
 \caption{\basic}
 \begin{algorithmic}[1]
 \setcounter{ALC@unique}{0}
 \renewcommand{\algorithmicrequire}{\textbf{Input:}}
 \renewcommand{\algorithmicensure}{\textbf{Output:}}
 \REQUIRE target state set $\mathcal{X}_T$, trained NN control policy $\pi$, time horizon $\tau$, partition parameter $\mathbf{r}$
 \ENSURE BP set approximations $\bar{\mathcal{P}}_{-\tau:0}(\mathcal{X}_T)$,
 affine control bound parameters
 $\mathbf{\Omega}_{-\tau:-1}$
    \STATE $\bar{\mathcal{P}}_{0}(\mathcal{X}_T) \leftarrow \mathcal{X}_T$ \label{alg:backprojection:initialize}
    \FOR{$t$ in $\{-1, -2, \ldots, -\tau\}$} \label{alg:backprojection:timestep_for_loop}
        \STATE $\bar{\bar{\mathcal{P}}}_{t+1}(\mathcal{X}_T) \leftarrow \text{boundWithRectangle}(\bar{\mathcal{P}}_{t+1}(\mathcal{X}_T))$ \label{alg:backprojection:bound_with_rect}
        \STATE $\bar{\mathcal{P}}_{t}(\mathcal{X}_T) \leftarrow \text{oneStepBackproj}(\bar{\bar{\mathcal{P}}}_{t+1}(\mathcal{X}_T), \pi, \mathbf{r})$ \label{alg:backprojection:one_step_backproj_iter}
        \STATE $[\ubar{\mathbf{x}}'_{t}, \bar{\mathbf{x}}'_{t}] \leftarrow \bar{\mathcal{P}}_{t}(\mathcal{X}_T) $
        \STATE $\mathbf{\Omega}_{t} = [\bar{\CROWNAu}, \bar{\CROWNAl}, \bar{\CROWNbu}, \bar{\CROWNbl}] \leftarrow \mathrm{CROWN}(\pi, [\ubar{\mathbf{x}}'_{t}, \bar{\mathbf{x}}'_{t}])$
    \ENDFOR
 \RETURN $\bar{\mathcal{P}}_{-\tau:0}(\mathcal{X}_T),\ \mathbf{\Omega}_{-\tau:-1}$ \label{alg:backprojection:return}
 \end{algorithmic}\label{alg:backprojection}
\end{algorithm}

\subsection{Algorithm for Computing N-Step Backprojection Sets}
Notice that by iteratively making over-approximations using the previously calculated BP over-approximation, \basic{} tends to accrue conservativeness over the time horizon due to the wrapping effect \cite{le2009reachability}.
Thus even if $\bar{\mathcal{P}}_{-\tau}(\mathcal{X}_T)$ tightly bounds the set of states that reach $\bar{\mathcal{P}}_{-\tau+1}(\mathcal{X}_T)$, it may be an overly conservative estimate of the set of states that ultimately end up in $\mathcal{X}_T$ in $\tau$ timesteps.
To reduce the accrued conservativeness, we present \nstep{} (\cref{alg:Nbackprojection}) that uses \basic{} to initialize $\bar{\mathcal{P}}_{-\tau:0}(\mathcal{X}_T)$ and the collected affine control bound parameters $\mathbf{\Omega}_{-\tau:-1}$. 
We then use a procedure similar to the one outlined in \cref{sec:overapprox_one_step_BP}, but now we relax the NN over $\bar{\mathcal{P}}_t$ instead of $\bar{\mathcal{R}}_t$ and include additional constraints that require the future states of the system to progress through the future BP estimates and eventually reach the target set while satisfying the relaxed affine control bounds at each step along the way.
The number of LPs calculated in \cref{alg:Nbackprojection} is $N_{LP} = 2 n_x N_\mathbf{r} (2 \tau - 1)$, which is again linear in state dimension, but with the same issue given by $N_\mathbf{r}$.

\begin{lemma}\label[lemma]{thm:Nbackprojection}

Given an $m$-layer NN control policy $\pi:\R^{n_x}\to\R^{n_u}$, closed-loop dynamics $f: \R^{n_x} \times \Pi \to \R^{n_x}$ as in~\cref{eqn:ltv_dynamics,eqn:closed_loop_dynamics}, a set of BP estimates $\bar{\mathcal{P}}_{-\tau:0}(\mathcal{X}_T)$, a corresponding set of affine control bounds $\mathbf{\Omega}_{-\tau:-1}$, and target set $\mathcal{X}_T$, the following relations hold:
\begin{equation*}
    \mathcal{P}_{t}(\mathcal{X}_T) \subseteq \bar{\mathcal{P}}'_t(\mathcal{X}_T) \subseteq \bar{\mathcal{P}}_t(\mathcal{X}_T), \forall t \in \underbrace{\{\text{-}\tau, \text{-}\tau+1, \ldots, -1\}}_{\triangleq \mathcal{T}},
\end{equation*}
where $\bar{\mathcal{P}}'_t(\mathcal{X}_T) \triangleq \{\x_{t}\ |\ \ubar{\mathbf{x}}'_{t} \leq \mathbf{x}_{t} \leq \bar{\mathbf{x}}'_{t} \}$ with $\ubar{\mathbf{x}}'_{t}$ and $\bar{\mathbf{x}}'_{t}$ calculated using the LPs specified by \cref{eqn:Nbackprojection:lp_obj}.

\begin{proof}
Given dynamics from~\cref{eqn:ltv_dynamics,eqn:closed_loop_dynamics}, solve the following optimization problems for each state $\istate \in [n_x]$ and for each $t \in \mathcal{T}$,
\begin{equation}
\bar{\mathbf{x}}_{t; \istate} = \min_{\mathbf{x}_t, \mathbf{u}_t \in \mathcal{F}_{\bar{\mathcal{P}}'_t}} \mathbf{e}_k^\top \mathbf{x}_{t}, \quad \quad \ubar{\mathbf{x}}_{t; \istate} = \max_{\mathbf{x}_t, \mathbf{u}_t \in \mathcal{F}_{\bar{\mathcal{P}}'_t}} \mathbf{e}_k^\top \mathbf{x}_{t}, \label{eqn:Nbackprojection:lp_obj}
\end{equation}
where
\begin{equation}
\mathcal{F}_{\bar{\mathcal{P}}'_t} \! \triangleq \! \left\{
\! \mathbf{x}_{t}, \mathbf{u}_{t} \left| \ 
\begin{aligned}
& \mathbf{A} \mathbf{x}_t + \mathbf{B} \mathbf{u}_t + \mathbf{c} = \x_{t+1}. \\
& \x_{t+1} \in \bar{\mathcal{P}}_{t+1}, \\
& \pi^L_t(\x_{t}) \leq \mathbf{u}_{t} \leq \pi^U_t(\x_t),\\
& \x_t \in \bar{\mathcal{P}}_t(\mathcal{X}_T), 
\end{aligned}\right.\kern-\nulldelimiterspace 
\right\},
\label{eqn:Nbackprojection:lp_constr}
\end{equation}
with $\bar{\mathcal{P}}_0(\mathcal{X}_T) = \mathcal{X}_T$, and $\pi^L_t$ and $\pi^U_t$ obtained from $\mathbf{\Omega}_{-\tau:-1}$.

The final constraint in \cref{eqn:Nbackprojection:lp_constr} guarantees $\bar{\mathcal{P}}'_t(\mathcal{X}_T) \subseteq \bar{\mathcal{P}}_t(\mathcal{X}), \forall t \in \mathcal{T}$.
The third constraint ensures that the relations \cref{eqn:BPproof:rect_bound,eqn:BPproof:true_relaxed_BP,eqn:BPproof:true_BP} in the proof of \cref{thm:backprojection} holds for all $t \in \mathcal{T}$, thus guaranteeing $\mathcal{P}_{t}(\mathcal{X}_T) \subseteq \bar{\mathcal{P}}'_t(\mathcal{X}_T)$. It follows that $\mathcal{P}_{t}(\mathcal{X}_T) \subseteq \bar{\mathcal{P}}'_t(\mathcal{X}_T) \subseteq \bar{\mathcal{P}}_t(\mathcal{X}), \forall t \in \mathcal{T}$.
\end{proof}
\end{lemma}

Notice that the first two constraints provide the key advantage of the LPs solved by \cref{eqn:Nbackprojection:lp_obj} over those given by \cref{thm:backprojection} in that they require the state to trace back through the set of BPs leading to the original target set, thus providing a better approximation of the true BP set.

\begin{algorithm}[t]
 \caption{\nstep}
 \begin{algorithmic}[1]
 \setcounter{ALC@unique}{0}
 \renewcommand{\algorithmicrequire}{\textbf{Input:}}
 \renewcommand{\algorithmicensure}{\textbf{Output:}}
 \REQUIRE target state set $\mathcal{X}_T$, trained NN control policy $\pi$, time horizon $\tau$, partition parameter $\mathbf{r}$
 \ENSURE Refined BP set approximations $\bar{\mathcal{P}}'_{-\tau:0}(\mathcal{X}_T)$
    \STATE $\bar{\mathcal{P}}'_{-\tau:0}(\mathcal{X}_T) \leftarrow \emptyset$
    \STATE $\bar{\mathcal{P}}_{-\tau:0}(\mathcal{X}_T),\ \mathbf{\Omega}_{-\tau:-1} \leftarrow \! \text{\basic}(\mathcal{X}_T, \pi, \tau, \mathbf{r})$
    \STATE $\bar{\mathcal{P}}'_{-1:0}(\mathcal{X}_T) \leftarrow \bar{\mathcal{P}}_{-1:0}(\mathcal{X}_T)$
    \FOR{$t$ in $\{-2, \ldots, -\tau\}$} \label{alg:Nbackprojection:timestep_for_loop}
        \STATE $[\ubar{\ubar{\mathbf{x}}}_{t}, \bar{\bar{\mathbf{x}}}_{t}] \leftarrow \text{boundWithRectangle}(\bar{\mathcal{P}}_{t}(\mathcal{X}_T))$
        \STATE $\mathcal{S} \leftarrow \mathrm{partition}([\ubar{\ubar{\mathbf{x}}}_{t}, \bar{\bar{\mathbf{x}}}_{t}], \mathbf{r})$
        \FOR{$[\ubar{\mathbf{x}}_{t}, \bar{\mathbf{x}}_{t}]$ in $\mathcal{S}$}
            \STATE $\CROWNAu, \CROWNAl, \CROWNbu, \CROWNbl \leftarrow \mathrm{CROWN}(\pi, [\ubar{\mathbf{x}}_{t}, \bar{\mathbf{x}}_{t}])$
            \FOR{$\istate \in n_x$}
                \STATE $\bar{\bar{\mathbf{x}}}_{t;k} \leftarrow \! \mathrm{NStepLpMax}(\bar{\mathcal{P}}_{\text{-}\tau:0}, \mathbf{\Omega}_{\text{-}\tau:\text{-}1},\CROWNAu, \CROWNAl, \CROWNbu, \CROWNbl)$
                \STATE $\ubar{\ubar{\mathbf{x}}}_{t;k} \leftarrow \! \mathrm{NStepLpMin}(\bar{\mathcal{P}}_{\text{-}\tau:0},\ \mathbf{\Omega}_{\text{-}\tau:\text{-}1},\CROWNAu, \CROWNAl, \CROWNbu, \CROWNbl)$
            \ENDFOR
            \STATE $\mathcal{A} \leftarrow \{\mathbf{x}\ \lvert\ \forall \istate \in n_x,\ \ubar{\ubar{\mathbf{x}}}_{t;k} \leq \mathbf{x} \leq \bar{\bar{\mathbf{x}}}_{t;k}\}$
            \STATE $\bar{\mathcal{P}}'_{-1}(\mathcal{X}_T) \leftarrow \bar{\mathcal{P}}_{-1}(\mathcal{X}_T) \cup \mathcal{A}$ \label{alg:Nbackprojection:union}
        \ENDFOR
        \STATE $\bar{\mathcal{P}}'_t(\mathcal{X}_T) \leftarrow \bar{\mathcal{P}}'_{-1}(\mathcal{X}_T)$
    \ENDFOR
 \RETURN $\bar{\mathcal{P}}'_{-\tau:0}(\mathcal{X}_T)$ \label{alg:Nbackprojection:return}
 \end{algorithmic}\label{alg:Nbackprojection}
\end{algorithm}


\section{Numerical Results}
In this section we use numerical experiments to verify our algorithms and demonstrate their properties as they relate to each other and to the forward reachability tool proposed in \cite{everett2021reachability}.
First we show how \nstep{} can be used to reduce the conservativeness gathered by \basic{} and we quantify the additional computation cost.
We then show how \basic{} can be used in a collision-avoidance scenario that causes Reach-LP \cite{everett2021reachability} to fail.

All numerical results were collected with the LP solver {\tt cvxpy} \cite{diamond2016cvxpy} on a machine running Ubuntu 20.04 with an i7-6700K CPU and 32 GB of RAM.

\subsection{Double Integrator}

\begin{figure}[t]
\setlength\belowcaptionskip{-0.1\baselineskip}
\centering
\captionsetup[subfigure]{aboveskip=-1pt,belowskip=-1pt}
    \begin{subfigure}[t]{\columnwidth}
        \begin{tikzpicture}[fill=white]
            \node[anchor=south west,inner sep=0] (image) at (0,0) {\includegraphics[width=1\columnwidth]{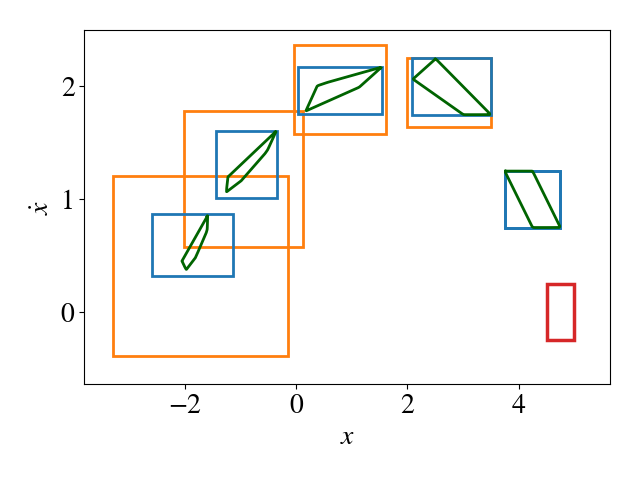}};
            \begin{scope}[x={(image.south east)},y={(image.north west)}]
                \node[] at (0.85,.25) {{\small Target Set}};
            \end{scope}
        \end{tikzpicture}
        \caption{BP set estimates and true BP convex hulls (dark green)}
        \label{fig:double_integrator_n_vs_one:BP_sets}
    \end{subfigure}
    \begin{subfigure}[t]{\columnwidth} 
        \begin{tikzpicture}[fill=white]
            \node[anchor=south west,inner sep=0] (image) at (0,0) {\includegraphics[width=1\columnwidth]{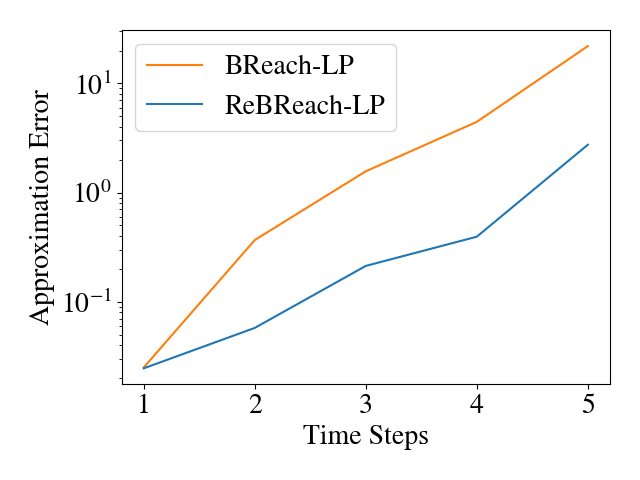}};
            
        \end{tikzpicture}
        \caption{Approximation error \cref{eqn:BP_estimate_error} calculated at each time step}
        \label{fig:double_integrator_n_vs_one:error_comparison}
    \end{subfigure}
    \caption{Compare BP set estimates for a double integrator extending from the target set (red) calculated with \basic{} (orange) and \nstep{} (blue).}
    \label{fig:double_integrator_n_vs_one}
\end{figure}

Consider the discrete-time double integrator model \cite{hu2020reach}
\begin{equation}
    \mathbf{x}_{t+1} =
    \underbrace{
    \begin{bmatrix}
    1 & 1 \\
    0 & 1
    \end{bmatrix}}_{\mathbf{A}} \mathbf{x}_t +
    \underbrace{
    \begin{bmatrix}
    0.5 \\ 1
    \end{bmatrix}}_{\mathbf{B}} \mathbf{u}_t
\end{equation}
with $\mathbf{c}=0$, $\mathbf{C}=\mathbf{I}_2$, and discrete sampling time $t_s=1$s. 
The NN controller (identical to the double integrator controller used in \cite{everett2021reachability}) has $[5,5]$ neurons, ReLU activations and was trained with state-action pairs generated by an MPC controller. 
\cref{fig:double_integrator_n_vs_one}  compares \basic{} (orange) and \nstep{} (blue). 
As shown, both algorithms collect some approximation error
\begin{equation}
    \mathrm{error} = \frac{A_\mathrm{true}-A_{BPE}}{A_\mathrm{true}},
    \label{eqn:BP_estimate_error}
\end{equation}
where $A_\mathrm{true}$ denotes the area of the tightest rectangular bound of the true BP set (dark green in~\cref{fig:double_integrator_n_vs_one:BP_sets}), calculated using Monte Carlo simulations, and $A_\mathrm{BPE}$ denotes the area of the BP estimate.
However, because of the additional constraints and partitioning steps included in~\nstep{}, it is able to reduce conservativeness in the final BP set estimate by 88\%, as shown in \cref{tab:alg_error_comparison}. 

\begin{table}[t]
\centering
\caption{Compare error~\cref{eqn:BP_estimate_error} for \basic{} and \nstep{} (reduces  conservativeness in final BP set estimate by 88\% with $2.5 \times$ computation of~\basic{}).}
\begin{tabular}{llr}
\hline
 Algorithm   & Runtime [s]       &   Final Step Error \\
\hline
 BReach-LP   & $1.349 \pm 0.046$ &              21.96 \\
 ReBReach-LP & $3.383 \pm 0.095$ &               2.74 \\
\hline
\end{tabular}
\label{tab:alg_error_comparison}
\vspace*{-0.12in}
\end{table}

\subsection{Linearized Ground Robot}
\label{sec:lgr}
\begin{figure*}[t]
\centering
 \captionsetup[subfigure]{aboveskip=-4pt,belowskip=-4pt}
        \begin{subfigure}[t]{0.325\textwidth}
        \begin{tikzpicture}[fill=white]
            \node[anchor=south west,inner sep=0] (image) at (0,0) {\includegraphics[width=\columnwidth]{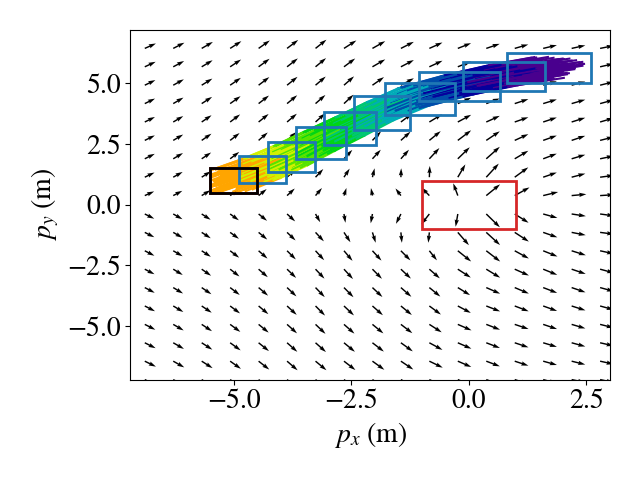}};
            \begin{scope}[x={(image.south east)},y={(image.north west)}]
                \node[draw,fill] at (0.44,0.3) {{\tiny $\mathcal{X}_0\! =\! \mathcal{B}_\infty \! \left( \begin{bmatrix} \text{-}5 \\ {\color{teal}1} \end{bmatrix}\!,\! \begin{bmatrix} 0.5 \\ 0.5 \end{bmatrix} \right)$}};
            \end{scope}
            \begin{scope}[x={(image.south east)},y={(image.north west)}]
                \node[] at (0.58,1) {{\color{teal}\small Certified Safe \ding{51}}};
            \end{scope}
        \end{tikzpicture}
        \caption{Nominal forward reachability collision avoidance scenario with vector field representation of control input. No intersection of target set (red) and reachable sets (blue) implies that safety can correctly be certified.}
        \label{fig:forward_reach_nominal}
    \end{subfigure}~
    \begin{subfigure}[t]{0.325\textwidth}
        \begin{tikzpicture}[fill=white]
            \node[anchor=south west,inner sep=0] (image) at (0,0) {\includegraphics[width=\columnwidth]{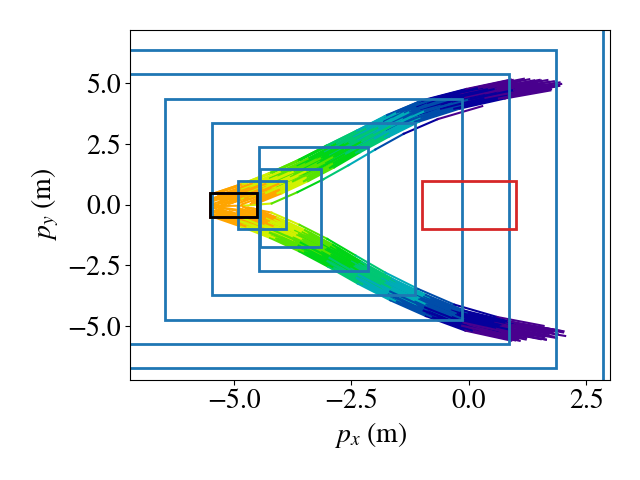}};
            \begin{scope}[x={(image.south east)},y={(image.north west)}]
                \node[draw,fill] at (0.44,0.3) {{\tiny $\mathcal{X}_0\! =\! \mathcal{B}_\infty \! \left( \begin{bmatrix} \text{-}5 \\ {\color{red}0} \end{bmatrix}\!,\! \begin{bmatrix} 0.5 \\ 0.5 \end{bmatrix} \right)$}};
                \node[text width=3cm] at (0.60,0.645) {{\tiny $t\text{=}1$}};
                \node[text width=3cm] at (0.675,0.67) {{\tiny $t\text{=}2$}};
                \node[text width=3cm] at (0.72,0.715) {{\tiny $t\text{=}3$}};
                \node[text width=3cm] at (0.78,0.765) {{\tiny $t\text{=}4$}};
                \node[text width=3cm] at (0.9,0.82) {{\tiny $t\text{=}5$}};
                \node[text width=3cm] at (0.99,0.865) {{\tiny $t\text{=}6$}};
            \end{scope}
            \begin{scope}[x={(image.south east)},y={(image.north west)}]
                \node[] at (0.58,1) {{\color{red}\small Possible Collision Detected \ding{55}}};
            \end{scope}
        \end{tikzpicture}
        \caption{Forward reachability strategy for collision avoidance at decision boundary. Reachable sets explode in response to uncertainty in which set of trajectories will be taken, causing an incorrect assessment of unsafe.}
         \label{fig:forward_reach_bifurcating}
    \end{subfigure}~
    \begin{subfigure}[t]{0.325\textwidth}
        \begin{tikzpicture}[fill=white]
            \node[anchor=south west,inner sep=0] (image) at (0,0) {\includegraphics[width=\columnwidth]{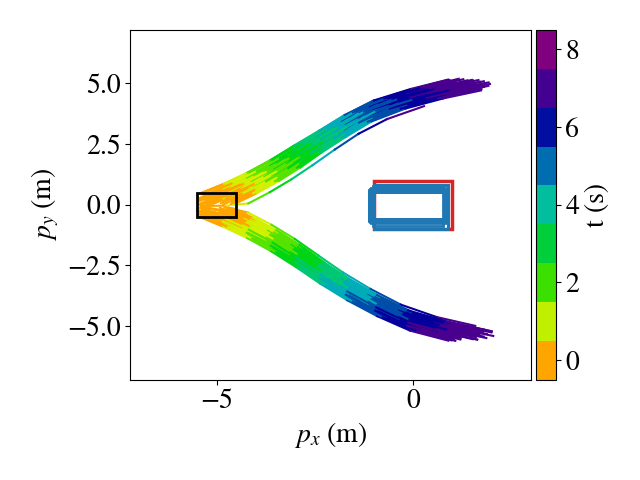}};
            \begin{scope}[x={(image.south east)},y={(image.north west)}]
                \node[draw,fill] at (0.44,0.3) {{\tiny $\mathcal{X}_0\! =\! \mathcal{B}_\infty \! \left( \begin{bmatrix} \text{-}5 \\ 0 \end{bmatrix}\!,\! \begin{bmatrix} 0.5 \\ 0.5 \end{bmatrix} \right)$}};
            \end{scope}
            \begin{scope}[x={(image.south east)},y={(image.north west)}]
                \node[] at (0.53,1) {{\color{teal}\small Certified Safe \ding{51}}};
            \end{scope}
            \begin{scope}[x={(image.south east)},y={(image.north west)}]
                \node[] at (0.64,.65) {{\tiny Target Set}};
            \end{scope}
        \end{tikzpicture}
        \caption{Backward reachability strategy for collision avoidance at decision boundary. Safety can correctly be certified because none of the BP set estimates (blue) intersect with the initial state set (black).}
        \label{fig:backward_reach_single_integrator}
    \end{subfigure}
    \caption{Collision-avoidance situation that \cite{everett2021reachability} incorrectly labels as dangerous whereas \basic{} correctly certifies safety.}
    \label{fig:lgr_comparison}
    \vspace*{-0.12in}
\end{figure*}
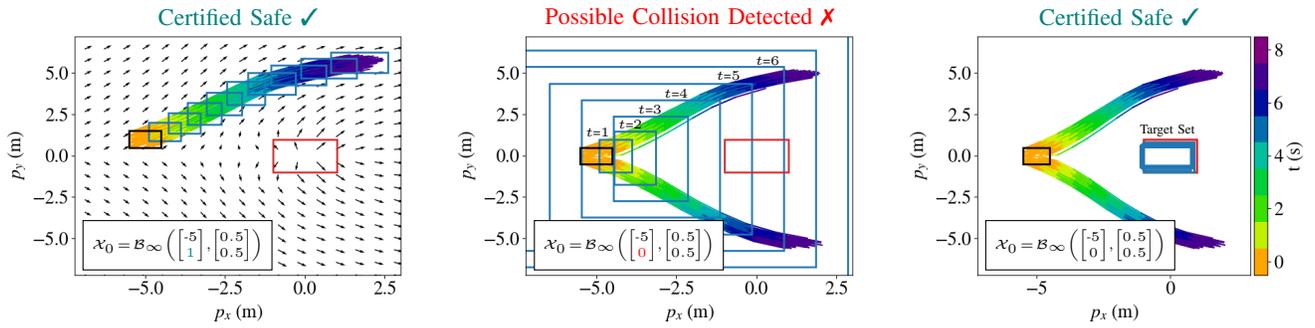

Using the feedback linearization technique proposed in \cite{martinez2021formation}, we represent the common unicycle model as a pair of integrators
\begin{equation}
    \mathbf{x}_{t+1} =
    \underbrace{
    \begin{bmatrix}
    1 & 0 \\
    0 & 1
    \end{bmatrix}}_{\mathbf{A}} \mathbf{x}_t +
    \underbrace{
    \begin{bmatrix}
    1 & 0 \\ 0 & 1
    \end{bmatrix}}_{\mathbf{B}} \mathbf{u}_t
\end{equation}
with $\mathbf{c}=0$, $\mathbf{C}=\mathbf{I}_2$, and sampling time $t_s=1$s.
With this system formulation, we can consider $\mathbf{x}_t=[p_x,p_y]^\top$ to represent the position of a vehicle in the $x\text{-}y$ plane and $\mathbf{u}_t=[v_x,v_y]^\top$.

To emulate the scenarios demonstrated by \cref{fig:reach_comp}, we trained a NN with [10,10] neurons and ReLU activations to mimic the vector field given by
\begin{equation}
    \mathbf{u}(\mathbf{x})\text{=}\!\!
    \begin{bmatrix} \!
    \mathrm{max}(\mathrm{min}(1+\frac{2p_x}{p_x^2+p_y^2},1),\text{-}1) \\
    \mathrm{max}(\mathrm{min}(\frac{p_y}{p_x^2+p_y^2}\!+\!2\mathfrak{s}(p_y)\frac{e^{-\frac{p_x}{2}+2}}{(1+e^{-\frac{p_x}{2}+2})^2},\!1),\!\text{-}1) \label{eqn:vector_field}\!
    \end{bmatrix}
\end{equation}
where $\mathfrak{s}(\cdot)$ returns the sign of the argument.
The vector field \cref{eqn:vector_field} is visualized in \cref{fig:forward_reach_nominal} and produces trajectories that drive the system away from an obstacle bounded by the target set $\mathcal{X}_T = \mathcal{B}_{\infty}([0,0]^\top,[1,1]^\top)$ (shown in red).
Eq.~\cref{eqn:vector_field} was used to generate $10^5$ data points sampled from the state space region $\mathcal{B}_{\infty}([0,0]^\top,[10,10]^\top)$, which were then used to train the NN for 20 epochs with a batch size of 32.

First, in \cref{fig:forward_reach_nominal}, we demonstrate a typical forward reachability example using the method described in \cite{everett2021reachability}.
The blue bounding boxes represent the forward reachable set estimates calculated using Reach-LP \cite{everett2021reachability} with $\mathbf{r} = [4,4]$ and the lines represent the time progression ($\mathbf{x}_0 \rightarrow \mathbf{x}_{\tau}: \mathrm{orange} \rightarrow \mathrm{purple}$) of a set of possible trajectories.
In this scenario, the system's initial state set $\mathcal{X}_0$ lies above the $x$-axis and the NN control policy uniformly commands the system to go above the obstacle. 
The resulting reachability analysis works as expected with reachable set estimates that tightly bound the future trajectories.

The scenario shown in \cref{fig:forward_reach_bifurcating} is identical to that of \cref{fig:forward_reach_nominal}, except the initial state is now centered on the $x$-axis. 
This scenario demonstrates a breakdown in standard reachability analysis tools due to the uncertainty in which trajectory will be taken by the system.
While some other tools, e.g., \cite{sidrane2021overt}, can be shown to reduce conservativeness in the upper and lower bounds of the reachable set estimates, the authors are not aware of any methods to remove the regions between the trajectories and correctly certify this situation as safe.
Also note that while it may initially seem like a good strategy to simply partition the initial set so that each element goes in one of the two directions, this would only work if the initial set could be split perfectly along the decision boundary, which may be difficult.
This is the also the reason that we solve LPs to find $\bar{\mathcal{P}}_t$ rather than propagate states from $\bar{\mathcal{R}}_t$ to $\mathcal{X}_T$ using forward reachability analysis.

Finally, in \cref{fig:backward_reach_single_integrator}, we demonstrate the same situation as in \cref{fig:forward_reach_bifurcating}, but now use backward reachability analysis as the strategy for safety certification.
Rather than propagating forward from the initial state set, we propagate backward from the target set, which was selected to bound the obstacle.
Because the control policy was designed to avoid the obstacle, the BP over-approximations, calculated using \basic{} with $\mathbf{r} = [4,4]$, do not intersect with the initial state set (black), thus implying that safety can be certified over the time horizon.
While the individual BP sets are harder to distinguish than the forward sets shown in \cref{fig:forward_reach_nominal,fig:forward_reach_bifurcating}, we use $\tau=9$ in each scenario, thereby checking safety over the same time horizon.
Note that the reachable sets in \cref{fig:forward_reach_bifurcating} were calculated in 0.5s compared to 2.35s for the BPs in \cref{fig:backward_reach_single_integrator}, but the result from \basic{} (\cref{fig:backward_reach_single_integrator}) provides more useful information.

\begin{figure}[t]
    \centering
    \begin{tikzpicture}[fill=white]
        \node[anchor=south west,inner sep=0] (image) at (0,0) {\includegraphics[width=1\columnwidth]{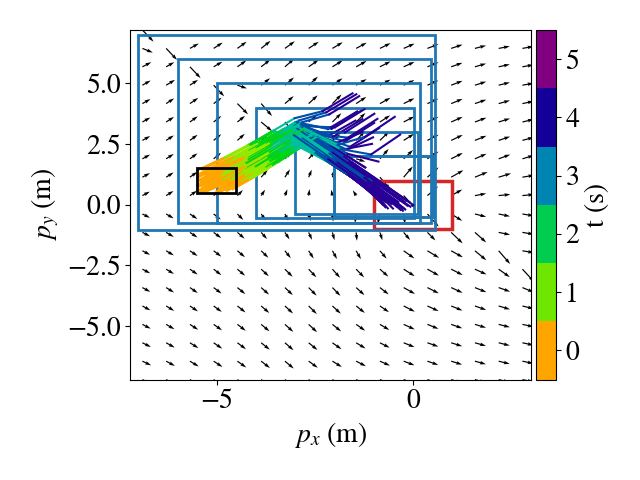}};
        \begin{scope}[x={(image.south east)},y={(image.north west)}]
            \node[draw,fill] at (0.4,0.3) {{\footnotesize $\mathcal{X}_0\! =\! \mathcal{B}_\infty \! \left( \begin{bmatrix} \text{-}5 \\ 1 \end{bmatrix}\!,\! \begin{bmatrix} 0.5 \\ 0.5 \end{bmatrix} \right)$}};
        \end{scope}
        \begin{scope}[x={(image.south east)},y={(image.north west)}]
            \node[] at (0.55,0.97) {{\color{red}\footnotesize Possible Collision Detected \ding{51}}};
        \end{scope}
    \end{tikzpicture} \vspace*{-0.4in}
    \caption{As expected, \nstep{} is unable to certify safety for a faulty NN control policy.}
    \label{fig:buggy_NN_demo}
    \vspace*{-0.18in}
\end{figure}

Finally, in \cref{fig:buggy_NN_demo} we confirm that our algorithms (in this case, \nstep{}) are able to detect a possible collision. Here we retrained the policy used in \cref{fig:lgr_comparison} but simulate a bug in the NN training process by commanding states along the line $y=-x$ to direct the system towards the obstacle at the origin. The initial state set is the same as that in \cref{fig:forward_reach_nominal}, but now we see that the system reaches the target set in 6 seconds. Because the 5th and 6th BP set estimates intersect with $\mathcal{X}_0$, \nstep{} cannot certify that the system is safe, thus demonstrating the desired behavior for a safety certification algorithm given a faulty controller.

\section{Conclusion}

This paper presented two algorithms, \basic{} and \nstep{}, for computing BP set estimates, i.e., sets for which a system will be driven to a designated target set, for linear NFLs over a given time horizon.
Because backward reachability analysis is challenging for systems with NN components, this work employs available forward analysis tools in a way that provides over-approximations of BP sets.
The key idea is to constrain the possible inputs of the system using typical analysis tools, then solve a set of LPs maximizing the size of the BP set subject to those constraints.
This technique is used iteratively by \basic{} to find BP set estimates multiple timesteps from the target set.
\nstep{} builds on \basic{} to include additional computations that reduce the conservativeness in the over-approximation.
Finally, we compared the performance of our two algorithms, demonstrating the trade-off between conservativeness and computation time, on a double integrator model, and compared our strategy to forward reachability in a collision-avoidance scenario with a linearized ground robot model.

Future work includes extending our methods to nonlinear NFLs, allowing us to handle more complex system dynamics.
Additionally, making use of symbolic propagation techniques inspired by \cite{sidrane2021overt} may allow for additional reductions in conservativeness of the BP set estimates.
Computation time may also be reduced by considering more efficient partitioning methods.

\balance
\bibliographystyle{IEEEtran}
\bibliography{refs}

\end{document}